\documentclass[aps,prd,reprint,floatfix]{revtex4}

\usepackage{amsfonts}
\usepackage{amssymb}
\usepackage{amsmath}
\usepackage{graphicx}
\usepackage{xcolor}
\usepackage[colorlinks=true,linkcolor=blue,citecolor=blue,urlcolor=blue]{hyperref}
\newcommand{\revision}[1]{\textcolor{black}{#1}}

\begin{document}

\title{Einstein-non-linear Maxwell-Yukawa black hole}
\author{S. Habib Mazharimousavi}
\email{habib.mazhari@emu.edu.tr}
\author{M. Halilsoy}
\email{mustafa.halilsoy@emu.edu.tr}
\affiliation{Department of Physics, Faculty of Arts and Sciences, Eastern Mediterranean
University, Famagusta, North Cyprus via Mersin 10, Turkey}
\date{\today }

\begin{abstract}
Within the context of nonlinear electromagnetism, we consider the Yukawa
extension of a Reissner-Nordstr\"{o}m black hole. An exact solution is given
which modifies certain characteristics of the latter. Some thermodynamical
aspects are given for comparison. \revision{The model provides a gauge-invariant nonlinear-electrodynamic realization of a screened electric potential; the Yukawa form by itself does not establish a massive vector excitation. The exact spacetime solution and its thermodynamic properties are examined with the exponential-integral contribution.}
\end{abstract}

\maketitle

\section{Introduction}

In $3+1-$dimensional static, spherically symmetric spacetimes, the solution
for the Einstein-Maxwell equations is known to be uniquely the Reissner-Nordstr
\"{o}m (RN) metric. Depending on the relative magnitude of mass and charge, 
the RN spacetime can be a black hole or a time-like naked singular geometry.
Extension of linear Maxwell to the nonlinear electrodynamics advocated first
by Born and Infeld \cite{1,2,3,4,5,6} has been considered extensively during
the recent decades. It is well-known that the original idea of Born-Infeld
was to eliminate Coulomb divergences for point charges. Elimination of
the spacetime divergence becomes the next step of expectation whenever
gravity is coupled to such a nonlinear electrodynamics. In this study we
consider a Yukawa extension of the Coulomb potential described by the scalar
potential, $\phi \left( r\right) =\frac{q}{r}e^{-\alpha r}$, where $q$ is
the electric charge and $\alpha $ is a positive constant. By this ansatz, the
theory involves two parameters, the charge $q$ and the Yukawa charge $\alpha$. The original idea of Yukawa \cite{7,8,9} which was introduced in
connection with nuclear interaction, irrespective of the electric charge,
was to provide room for massive mesons, satisfying a Proca-type equation $%
\left( \square -\alpha ^{2}\right) \phi \left( t,\mathbf{r}\right) =0.$ The $%
\square $ operator is to be defined on a flat spacetime which admits plane
wave solutions representing massive bosons/mesons. \ With the advent of
notions such as massive gravity and astrophysical objects dominated by
nuclear material, \revision{a Yukawa-type screening profile is a useful phenomenological ansatz. In the present gauge-invariant $\mathcal L(\mathcal F)$ model, however, the prescribed screened potential does not by itself imply a Proca mass or a massive electromagnetic mediator.} Our aim in the present study will be
restricted to static and spherically symmetric metrics in which a Yukawa
term is coupled with the standard Coulomb potential within the context of
nonlinear electrodynamics. The combination of two nonlinearities, as
expected, makes the theory highly nonlinear for an analytic treatment. For
instance, the nonlinear electromagnetic Lagrangian happens to be expressible
in terms of the radial coordinate explicitly, which, however, is
transcendental in terms of the field tensor. As a result, the model, its
basic features, and the roots of the metric function can only be analyzed
numerically. In spite of all such technical problems, we are able to identify
black hole solutions in such a theory. Not to mention, the origin $r=0$ is a
singular point of both the Maxwell field as well as the spacetime. There are
cases that admit double/single roots or no roots at all to define black
holes and naked singularities, respectively. Asymptotic behaviors at near ($%
r\rightarrow 0$) and far ($r\rightarrow \infty $) distances can be found
easily. The role of the Yukawa term, i.e., the parameter $\alpha,$ can be
described without much effort, and it serves to suppress the electric field exponentially at large radii. It is observed that by a scaling of the radial
coordinate the Yukawa parameter can be washed out so that the general
behavior of the solution is universal, irrespective of $\alpha$.

Let us add that the Yukawa gravitational potential has already been
considered in the literature. For instance, in the framework of $f\left(
R\right) $ modified theory of gravity, I. De Martino et al. very recently have
studied the Yukawa gravitational potential \cite{10,11}. Yukawa-type
potential in a cosmological model was also considered by M. O. Ribas et al. in 
\cite{12,13}. Also, a Yukawa-like potential in elliptical galaxies within the 
$f\left( R\right) $ gravity was considered earlier in \cite{14} by N. R.
Napolitano et al. The effect of such a kind of potential on the perihelion
precession of bodies in the solar system has been studied in \cite{15}.
There are other works which consider the Yukawa correction to the standard
gravity, which can be seen in \cite{16,17,18,19}. But to the best of our
knowledge, there has not been any study on gravity coupled with a
Yukawa-like electric potential. We note that although the original idea in
BI theory was to eliminate the singularity of the electric field, the recent
nonlinear electrodynamic theories do not necessarily obey this criterion 
\cite{20,21,22,23,24,25,26,27,28,29}. Furthermore, gravity coupled with
nonlinear electrodynamics has attracted intensive attention in the
literature \cite%
{30,31,32,33,34,35,36,37,38,39,40,41,42,43,44,45,46,47,48,49,50,51,52,53,54,55}%
. Adding all papers in this list is not possible, and what we referred to here
are part of the works which obviously show how important the subject is. To
complete our literature review, we would like to mention some works in this
field which have almost become an icon (perhaps in view of the authors of
the current work). The first paper is one of the earliest models of the
nonlinear electrodynamics proposed in 1936 by Heisenberg and Euler \cite%
{56,57}.The other two papers in this list are about the existence/nonexistence
of the regular black hole in gravity coupled with nonlinear electrodynamics
by E. Ay\'{o}n-Beato and A. Garc\'{\i}a, \cite{58} and K. A. Bronnikov \cite%
{59,60}. The next paper, which should be in our list, may be considered one of
the first papers in popularizing the BI theory \cite{61}. Finally, we would
like to mention two papers which applied the BI theory in more physical
manners \cite{62,63}.

The organization of the paper is as follows. In Section II we introduce the
Einstein-Nonlinear Maxwell-Yukawa (ENLMY) metric, derive field equations and
solve them. In Section III we study the physical properties of the solution
obtained in Section II. The paper is completed with our Conclusion in
Section IV.

\section{The Einstein-Nonlinear Maxwell-Yukawa metric}

Yukawa potential in the spherical coordinate system is given by%
\begin{equation}
\phi \left( r\right) =\frac{q}{r}e^{-\alpha r}
\end{equation}%
in which $q$ is the charge parameter and $\alpha $ is a positive constant. \revision{The conserved charge associated with the nonlinear electric displacement is fixed by the generalized Maxwell flux, rather than by the asymptotic flux of the exponentially suppressed electric field.} In the differential geometry formalism, we consider the
electric potential one-form to be given by%
\begin{equation}
\mathbf{A}=\phi \left( r\right) dt.
\end{equation}%
The Maxwell's field two-form is obtained to be%
\begin{equation}
\mathbf{F}=Edt\wedge dr
\end{equation}%
in which, $\wedge $ stands for the wedge product and%
\begin{equation}
E=-\phi ^{\prime }\left( r\right) =\frac{q\left( 1+\alpha r\right) }{%
r^{2}e^{\alpha r}}.
\end{equation}%
Considering the static and spherically symmetric line element in the form 
\begin{equation}
ds^{2}=-f\left( r\right) dt^{2}+\frac{dr^{2}}{f\left( r\right) }+r^{2}\left(
d\theta ^{2}+\sin ^{2}\theta d\varphi ^{2}\right) 
\end{equation}%
we find the Maxwell invariant $\mathcal{F}=F_{\mu \nu }F^{\mu \nu }$
determined as%
\begin{equation}
\mathcal{F}=-2\left( \frac{q\left( 1+\alpha r\right) }{r^{2}e^{\alpha r}}%
\right) ^{2}.
\end{equation}%
We consider now a general nonlinear electrodynamic Lagrangian of the form $%
\mathcal{L}=\mathcal{L}\left( \mathcal{F}\right) $ for the Maxwell-Yukawa
potential and write the Einstein-Nonlinear-Maxwell-Yukawa (ENLMY) action to
be of the form%
\begin{equation}
S=\frac{1}{16\pi G}\int d^{4}x\sqrt{-g}\left( R+\mathcal{L}\left( \mathcal{F}%
\right) \right) 
\end{equation}%
in which $R$ is the Ricci scalar. The variation of the action with respect
to $A_{\mu }$ yields the Maxwell-Yukawa equation given by%
\begin{equation}
d\left( \mathbf{\tilde{F}}\frac{\partial \mathcal{L}}{\partial \mathcal{F}}%
\right) =0
\end{equation}%
where $\mathbf{\tilde{F}}$ is the dual-Maxwell field given by the two-form 
\begin{equation}
\mathbf{\tilde{F}}=Er^{2}\sin \theta d\theta \wedge d\varphi .
\end{equation}%
This equation implies, within spherical symmetry, that
\begin{equation}
\frac{\partial \mathcal{L}}{\partial \mathcal{F}}=\frac{C_{0}e^{\alpha r}}{%
q\left( 1+\alpha r\right) }
\end{equation}%
in which $C_{0}$ is an integration constant. Next, we vary the action with
respect to $g_{\mu \nu }$ to find the ENLMY equations given by ($8\pi G=1$)%
\begin{equation}
G_{\mu }^{\nu }=T_{\mu }^{\nu }
\end{equation}%
in which 
\begin{equation}
T_{\mu }^{\nu }=\frac{1}{2}\left( \mathcal{L}\delta _{\mu }^{\nu }-4\frac{%
\partial \mathcal{L}}{\partial \mathcal{F}}F_{\mu \lambda }F^{\nu \lambda
}\right).
\end{equation}%
Explicitly, we find%
\begin{equation}
T_{t}^{t}=T_{r}^{r}=\frac{1}{2}\left( \mathcal{L}-2\mathcal{F}\frac{\partial 
\mathcal{L}}{\partial \mathcal{F}}\right) 
\end{equation}%
while%
\begin{equation}
T_{\theta }^{\theta }=T_{\varphi }^{\varphi }=\frac{1}{2}\mathcal{L}.
\end{equation}%
To have the ENLMY equation solved, we need to find the closed form of $%
\mathcal{L}$. To find this, we use the relation 
\begin{equation}
\frac{\partial \mathcal{L}}{\partial r}=\frac{\partial \mathcal{L}}{\partial 
\mathcal{F}}\frac{\partial \mathcal{F}}{\partial r}
\end{equation}%
which yields a first-order differential equation given by%
\begin{equation}
\frac{\partial \mathcal{L}}{\partial r}=\frac{4C_{0}q\left( 1+\left(
1+\alpha r\right) ^{2}\right) }{r^{5}e^{\alpha r}}.
\end{equation}%
The latter admits a solution of the form%
\begin{widetext}
\begin{equation}
\mathcal{L}=\frac{C_{0}q}{r^{4}}\left( \left( \alpha ^{3}r^{3}-1-\left(
1+\alpha r\right) ^{2}\right) e^{-\alpha r}-\alpha ^{4}r^{4}\mathcal{E}%
_{1}\left( \alpha r\right) \right) +C_{1}
\end{equation}
\end{widetext}%
in which $C_{1}$ is an integration constant, and the Exponential Integral is
defined to be \cite{64} 
\begin{equation}
\mathcal{E}_{1}\left( x\right) =\int_{1}^{\infty }\frac{e^{-xt}}{t}dt.
\end{equation}%
Let us note at this point that inversion of $r$ in terms of $\mathcal{F}$ is
a transcendental expression; for this reason, we shall make use of the
Lagrangian only implicitly as a function of $\mathcal{F}$. Nonzero $C_{1}$
may be considered as a cosmological constant, which we prefer not to take
into account in the present study. Hence, we set it to zero, i.e., $%
C_{1}=0.$ \revision{ For $C_0=-q$, the generalized electric displacement obeys $r^2\mathcal L_{\mathcal F}E=-q$. Thus $q$ is the conserved nonlinear electromagnetic charge in the adopted normalization, whereas $r^2E\to0$ as $r\to\infty$. Furthermore, $\mathcal L_{\mathcal F}=-e^{\alpha r}/(1+\alpha r)$ diverges as $r\to\infty$ at fixed $\alpha>0$; the ordinary Maxwell limit is obtained as $\alpha\to0$, not as a standard weak-field limit at fixed screening parameter.} 
\begin{figure}[tbp]
\includegraphics[width=120mm]{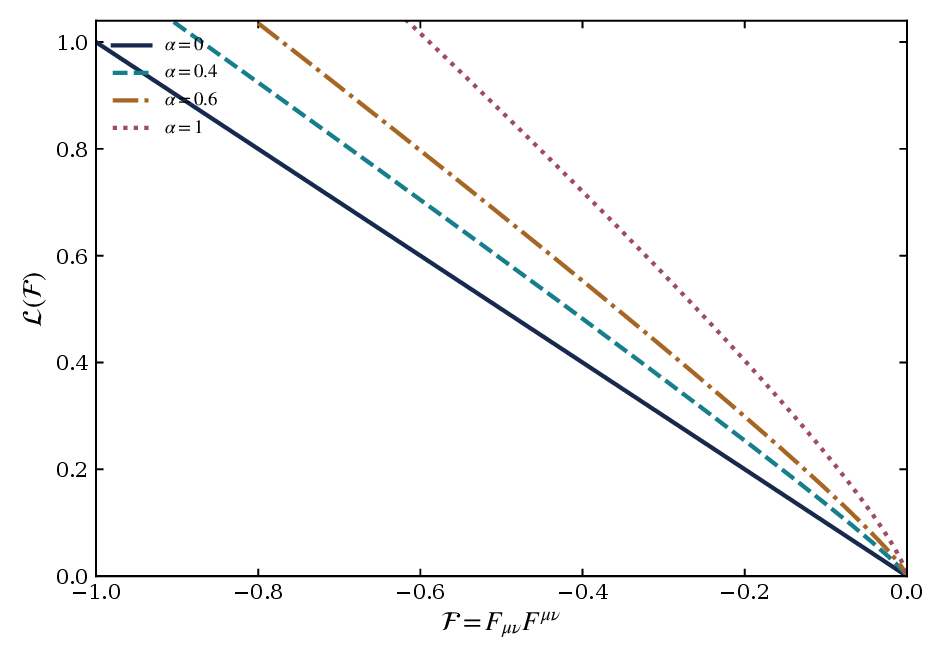}
\caption{{}$\mathcal{L}$ versus $\mathcal{F}$ for $\alpha %
=0.0,0.4,0.6$ and $1.0$ from left to right, respectively, and $q=1$. }
\end{figure}
In Fig. 1 we plot $\mathcal{L}$ versus $\mathcal{F}$ for some different
values of $\alpha $ and $q=1.$ The deviation from the linear Maxwell's
theory increases with increasing $\alpha .$ Having found $\mathcal{L},%
\mathcal{F}$ and $\frac{\partial \mathcal{L}}{\partial \mathcal{F}}$ in
terms of $r$ we are ready to express the solution for the ENLMY equations.
We obtain from (11) the metric function exactly to be%
\begin{widetext}
\begin{equation}
f\left( r\right) =1-\frac{2M}{r}+\frac{qC_{0}\left( \alpha ^{3}r^{3}-\alpha
^{2}r^{2}+2\alpha r-6\right) e^{-\alpha r}}{6r^{2}}-\frac{\alpha ^{4}}{6}%
r^{2}qC_{0}\mathcal{E}_{1}\left( \alpha r\right) .
\end{equation}
\end{widetext}%
Herein, $M$ is an integration constant to be interpreted as mass. \revision{A direct consistency check uses $G^t{}_t=(rf^{\prime}+f-1)/r^2$ and $G^\theta{}_\theta=f^{\prime\prime}/2+f^{\prime}/r$. Substitution of the metric above and the reconstructed nonlinear Lagrangian yields $G^t{}_t=T^t{}_t$ and $G^\theta{}_\theta=T^\theta{}_\theta$.} To
identify $C_{0}$ we expand the metric function for small $\alpha $ which
yields%
\begin{widetext}
\begin{equation}
f\left( r\right) \simeq 1-\frac{2M}{r}-\frac{qC_{0}}{r^{2}}+\frac{4qC_{0}}{3r%
}\alpha -qC_{0}\alpha ^{2}+\frac{2qC_{0}}{3}r\alpha ^{3}+\mathcal{O}\left(
\alpha ^{4}\right) .
\end{equation}
\end{widetext}%
In the limit $\alpha \rightarrow 0$ it should reproduce the Reissner-Nordstr\"om%
\"{o}m black hole solution which in turn yields $C_{0}=-q.$ The final form
of the ENLMY black hole solution can be expressed exactly as%
\begin{widetext}
\begin{equation}
f\left( r\right) =1-\frac{2M}{r}-\frac{q^{2}\left( \alpha ^{3}r^{3}-\alpha
^{2}r^{2}+2\alpha r-6\right) e^{-\alpha r}}{6r^{2}}\revision{+\frac{\alpha ^{4}}{6}%
r^{2}q^{2}\mathcal{E}_{1}\left( \alpha r\right) }.
\end{equation}
\end{widetext}%
\revision{The plus sign of the exponential-integral contribution follows directly from Eq.~(19) after setting $C_0=-q$ and is required by both independent Einstein equations.} For $\alpha \rightarrow 0$ this admits the expression%
\begin{widetext}
\begin{equation}
f_{\alpha \rightarrow 0}\left( r\right) \simeq 1-\frac{2M}{r}+\frac{q^{2}}{%
r^{2}}-\frac{4q^{2}}{3r}\alpha +q^{2}\alpha ^{2}-\frac{2q^{2}}{3}r\alpha
^{3}+\revision{\mathcal{O}\left( \alpha ^{4}\lvert\ln\alpha\rvert\right)} .
\end{equation}
\end{widetext}%
Note that introducing $\rho =\alpha r$ , $Q=\alpha q$ and $m=\alpha M$ the
metric function becomes%
\begin{equation}
f\left( \rho \right) =1-\frac{2m}{\rho }-\frac{Q^{2}\left( \rho ^{3}-\rho
^{2}+2\rho -6\right) e^{-\rho }}{6\rho ^{2}}\revision{+\frac{Q^{2}}{6}\rho ^{2}%
\mathcal{E}_{1}\left( \rho \right) } 
\end{equation}%
which shows that $\alpha $ acts as a scale factor so that it can be chosen
as $\alpha =1$. Hence, irrespective of the value of $\alpha ,$ one can study
the global properties of the resulting solution. \revision{The metric profile is plotted in Fig. 2 for $m=1$ and several values of $Q$, together with the Reissner-Nordstr\"om comparison in Fig. 3.} 
\begin{figure}[tbp]
\includegraphics[width=120mm]{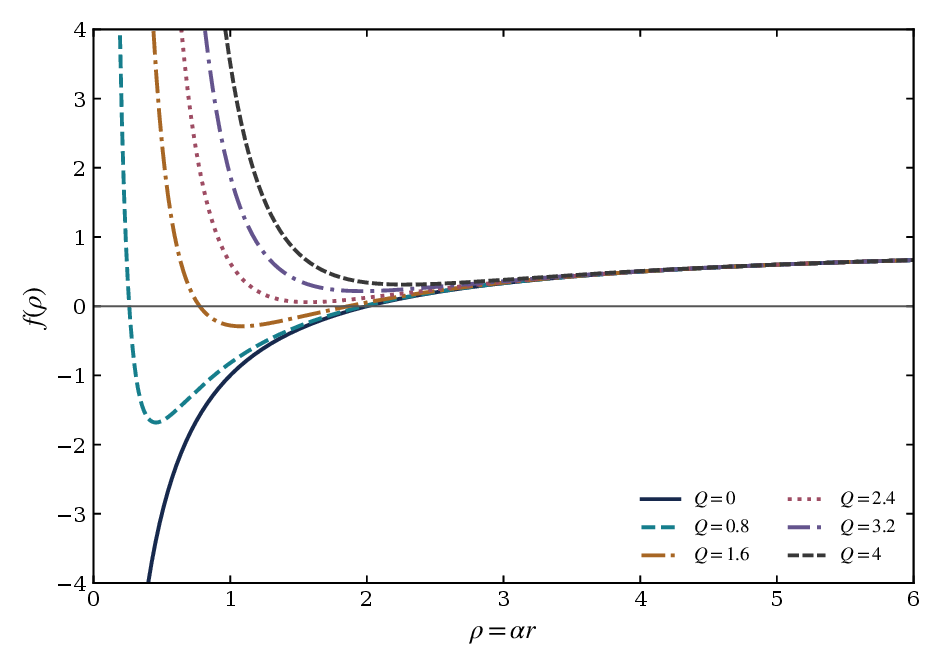}
\caption{\revision{Metric function $f(\rho)$ for $m=1$ and $Q=0,0.8,\ldots,4.0$. The intersections with the horizontal axis determine the horizons.}}
\end{figure}
\begin{figure}[tbp]
\includegraphics[width=120mm]{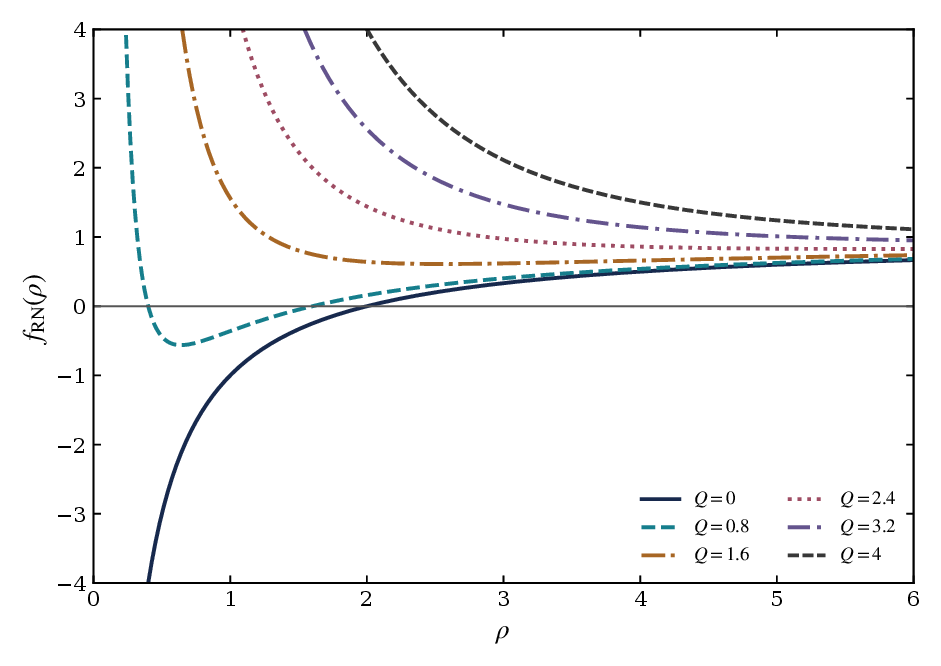}
\caption{{}Metric function for the Reissner-Nordstr\"{o}m spacetime $f\left( 
\protect\rho \right) =1-\frac{2m}{\protect\rho }+\frac{q^{2}}{\protect\rho %
^{2}}$ in terms of $\protect\rho $ with $m=1$ and from below $%
Q=0.0,0.8,...,4.0.$ The difference between RN and ENLMY metrics can be seen
by comparison between this figure and Fig. 2, which corresponds to the same
mass and charge.}
\end{figure}

\section{Physical properties of the ENLMY black hole}

The scaled metric function (23) is asymptotically flat and for small $\rho $
behaves as 
\begin{equation}
f_{\rho \rightarrow 0}\left( \rho \right) \simeq 1+Q^{2}-\frac{6m+4Q^{2}}{%
3\rho }+\frac{Q^{2}}{\rho ^{2}}-\frac{2Q^{2}}{3}\rho +\revision{\mathcal{O}\left( \rho^{2}\lvert\ln\rho\rvert\right)}
\end{equation}%
\revision{which gives $f(\rho)\to+\infty$ for $Q\neq0$. Accordingly, the geometry can describe two horizons, an extremal horizon, or no horizon, depending on $(m,Q)$. The extremal boundary is fixed by the simultaneous equations $f(\rho_e)=f'(\rho_e)=0$; the boundary is displayed in Fig. 4. In contrast with Reissner-Nordstr\"om spacetime, the boundary is not a straight line in the $(m,Q)$ plane.} 
\begin{figure}[tbp]
\includegraphics[width=120mm]{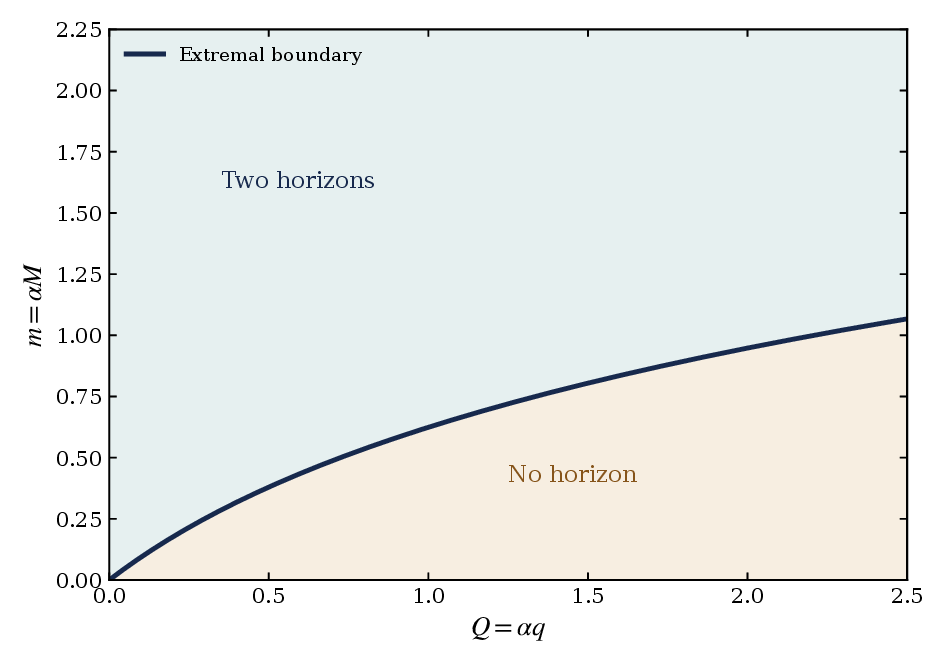}
\caption{\revision{Extremality curve in the $(Q,m)$ plane. The boundary is obtained by solving $f(\rho_e)=f\prime(\rho_e)=0$; points above it possess two positive roots and those below it possess none, for the parameter range shown.}}
\end{figure}
\begin{figure}[tbp]
\includegraphics[width=120mm]{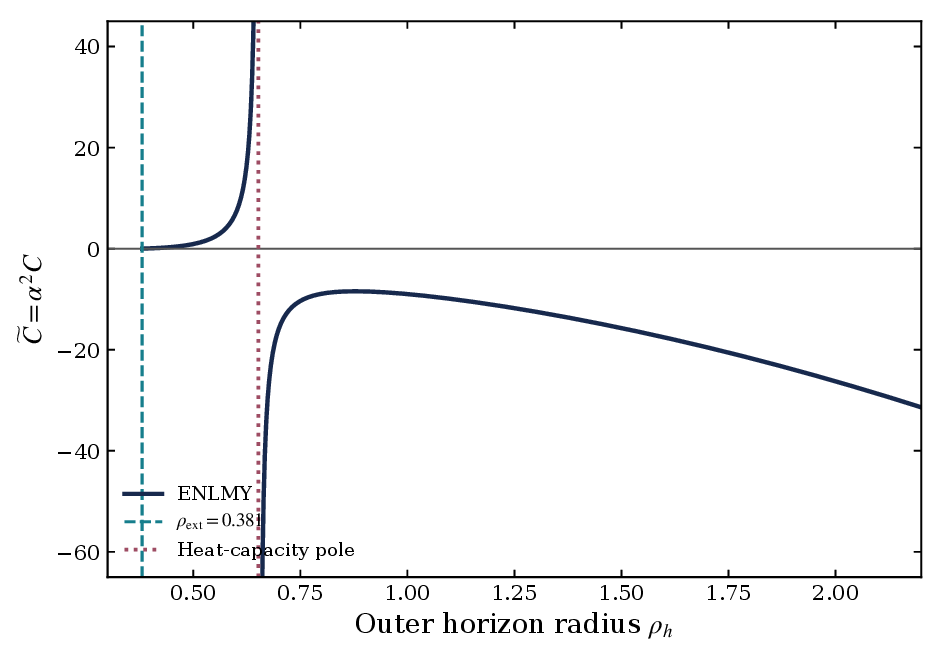}
\caption{\revision{Heat capacity on the physical outer-horizon branch for $Q=0.4$. The extremal radius is approximately $\rho_e=0.38057$; a heat-capacity pole occurs near $\rho_h=0.65208$. A pole alone does not establish a thermodynamic phase transition.}}
\end{figure}
\begin{figure}[tbp]
\includegraphics[width=120mm]{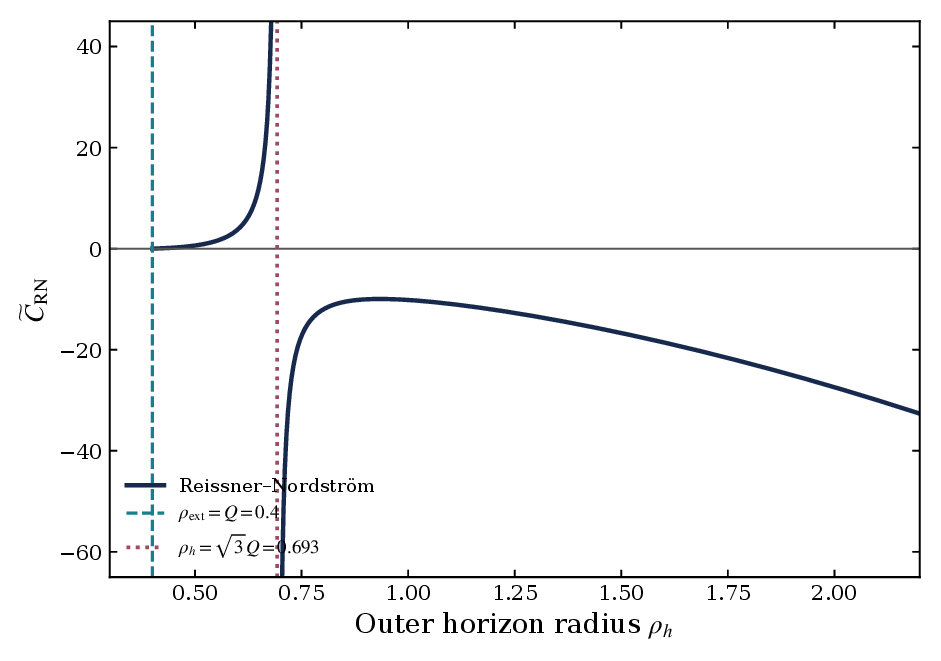}
\caption{{}Heat capacity of RN solution is depicted in terms of the
horizon's radius for $Q=0.4.$ We see from Fig. 5 that the heat capacity of
the ENLMY black hole for small $\protect\alpha $ is much smaller in
magnitude than its RN counterpart.}
\end{figure}
For the ENLMY black hole with an event horizon located at $r=r_{h}$ ($\rho
=\rho _{h}$) one finds the Hawking temperature 
\begin{equation}
T_{H}=\frac{1}{4\pi }\left. \frac{\partial f\left( r\right) }{\partial r}%
\right\vert _{r=r_{h}}.
\end{equation}%
After we introduce the scaled parameters and variables, the scaled Hawking
temperature is found to be%
\begin{widetext}
\begin{equation}
\tilde{T}_{H}=\frac{T_{H}}{\alpha }=\frac{1}{4\pi }\left. \frac{\partial
f\left( \rho \right) }{\partial \rho }\right\vert _{\rho =\rho _{h}}=\revision{\frac{
2\rho _{h}^{2}-Q^{2}\left( 2+2\rho _{h}-\rho _{h}^{2}+\rho _{h}^{3}\right)
e^{-\rho _{h}}+Q^{2}\rho _{h}^{4}\mathcal{E}_{1}\left( \rho _{h}\right) }{
8\pi \rho _{h}^{3}}}.
\end{equation}
\end{widetext}%
\revision{The zero-charge limit of this expression is $\tilde T_H=1/(4\pi\rho_h)$, as required for the Schwarzschild solution.} Furthermore, from the scaled entropy $\tilde{S}=\pi \rho _{h}^{2}$ one finds
the scaled heat capacity of the black hole given by%
\begin{widetext}
\begin{equation}
\tilde{C}=\alpha ^{2}C=\tilde{T}_{H}\left( \frac{\partial \tilde{S}}{%
\partial \tilde{T}_{H}}\right) _{Q}=\frac{2\pi \rho _{h}^{2}\left[ 2\rho
_{h}^{2}-Q^{2}\left( 2+2\rho _{h}-\rho _{h}^{2}+\rho _{h}^{3}\right)
e^{-\rho _{h}}+Q^{2}\rho _{h}^{4}\mathcal{E}_{1}\left( \rho _{h}\right) %
\right] }{-2\rho _{h}^{2}-Q^{2}\left( -6-6\rho _{h}-\rho _{h}^{2}+\rho
_{h}^{3}\right) e^{-\rho _{h}}+Q^{2}\rho _{h}^{4}\mathcal{E}_{1}\left( \rho
_{h}\right) }.
\end{equation}
\end{widetext}%
\revision{A zero of its numerator can correspond to an extremal configuration, whereas a zero of its denominator indicates a divergence of the canonical heat capacity. Such a divergence may signal a change of local thermodynamic stability, but does not by itself prove the existence of a phase transition. Figure 5 shows the heat capacity for $Q=0.4$ on the outer-horizon branch. The corresponding Reissner-Nordstr\"om result is presented in Fig. 6 for comparison.}

\section{Conclusion}

\revision{In this study, we considered the Yukawa modification of the Coulomb potential, $\phi(r)=q e^{-\alpha r}/r$, as a prescribed electric profile in Einstein gravity coupled to nonlinear electrodynamics. The nonlinear electromagnetic Lagrangian can be reconstructed parametrically in terms of the radial coordinate, and the corresponding Einstein equations yield an exact static and spherically symmetric solution. The exponential-integral term in the metric is essential for satisfying these equations. The geometry retains a central curvature singularity and, depending on the mass and charge parameters, admits black holes with two horizons, extremal black holes, or naked singularities.}

\revision{The Hawking temperature and canonical heat capacity can be determined analytically in terms of the horizon radius. In particular, the temperature has the required Schwarzschild limit, while the heat-capacity poles identify possible changes in local canonical stability. The Yukawa screening of the electric field does not by itself imply a Proca mass for the gauge field: the present construction is a gauge-invariant nonlinear-electrodynamic model and not a theory of an explicitly massive photon. Moreover, the standard Maxwell theory is recovered in the parameter limit $\alpha\to0$, rather than in the weak-field limit at fixed positive $\alpha$.}

\revision{The same reconstruction method can be applied to other prescribed electric profiles. An example is $\phi(r)=q e^{-\alpha r}/\sqrt{r^2+b^2}$, where $b$ is an additional length scale; its consistency and geometry would require a separate analysis.}

\end{document}